\documentclass[conference]{IEEEtran}
\IEEEoverridecommandlockouts

\usepackage{cite}
\usepackage{amsmath,amssymb,amsfonts}
\usepackage{algorithmic}
\usepackage{graphicx}
\usepackage{textcomp}
\usepackage{xcolor}
\usepackage{subcaption}
\usepackage{caption}  
\usepackage{titlesec}
\titlespacing{\section}{0.5pt}{0.5pt}{0.5pt}  
\titlespacing{\subsection}{0.5pt}{0.5pt}{0.5pt}

\def\BibTeX{{\rm B\kern-.05em{\sc i\kern-.025em b}\kern-.08em
    T\kern-.1667em\lower.7ex\hbox{E}\kern-.125emX}}
\begin{document}

\title{Dynamic Activation and Assignment of SDN Controllers in LEO Satellite Constellations}
\author{
\IEEEauthorblockN{1\textsuperscript{st} Wafa Hasanain, 2\textsuperscript{nd} Pablo G. Madoery, 3\textsuperscript{rd} Halim Yanikomeroglu}
\IEEEauthorblockA{
\textit{Non-Terrestrial Networks (NTN) Lab, Dept. of Systems and Computer Engineering} \\
\textit{Carleton University}, Ottawa, Canada
}
\and
\IEEEauthorblockN{}
\and
\IEEEauthorblockN{4\textsuperscript{th} Gunes Karabulut Kurt}
\IEEEauthorblockA{
\textit{Poly-Grames Research Center, Dept. of Electrical Engineering} \\
\textit{Polytechnique Montréal}, Montréal, Canada
}
\and
\IEEEauthorblockN{5\textsuperscript{th} Sameera Siddiqui}
\IEEEauthorblockA{
\textit{Defence Research and Development Canada}, \\Canada
}
\and
\IEEEauthorblockN{}
\and
\IEEEauthorblockN{6\textsuperscript{th} Stephane Martel}
\IEEEauthorblockA{
\textit{Satellite Systems, MDA}, Canada
}
\and
\IEEEauthorblockN{7\textsuperscript{th} Khaled Ahmed}
\IEEEauthorblockA{
\textit{Satellite Systems, MDA}, Canada
}
\and
\IEEEauthorblockN{8\textsuperscript{th} Colin Bellinger}
\IEEEauthorblockA{
\textit{National Research Council Canada}, Canada
}
}

\maketitle

\begin{abstract}
Software-defined networking (SDN) has emerged as a promising approach for managing traditional satellite communication. This enhances opportunities for future services, including integrating satellite and terrestrial networks. In this paper, we have developed an SDN-enabled framework for Low Earth Orbit (LEO) satellite networks, incorporating the OpenFlow protocol, all within an OMNeT++ simulation environment.
Dynamic controller assignment is one of the most significant challenges for large LEO constellations. Due to the movement of LEO satellites, satellite-controller assignments must be updated frequently to maintain low propagation delays. To address this issue, we present a dynamic satellite-to-controller assignment (DSCA) optimization problem that continuously adjusts these assignments. Our optimal DSCA (Opt-DSCA) approach minimizes propagation delay and optimizes the number of active controllers.
Our preliminary results demonstrate that the DSCA approach significantly outperforms the static satellite-to-controller assignment (SSCA) approach. While SSCA may perform better with more controllers, this scheme fails to adapt to satellite movements. Our DSCA approach consistently improves network efficiency by dynamically reassigning satellites based on propagation delays. Further, we found diminishing returns when the number of controllers is increased beyond a certain point, suggesting optimal performance with a limited number of controllers. Opt-DSCA lowers propagation delays and improves network performance by optimizing satellite assignments and reducing active controllers.
\end{abstract}


\section{Introduction}
\thispagestyle{empty}
Non-terrestrial networks (NTN), particularly low Earth orbit (LEO) satellite networks, present a promising solution for global internet access. As large-scale constellations are deployed to meet growing connectivity demands, traditional architectures face challenges in flexibility and seamless terrestrial integration \cite{ref9}. The rapid movement of LEO satellites leads to frequent topology changes and latency variations, necessitating adaptive control mechanisms for stable communication. Software-defined networking (SDN) improves adaptability by decoupling control and data planes \cite{ref22}. Although prior studies \cite{ref13, ref8, ref14} address adaptive controller placement, many depend on static or periodic reconfigurations that do not fully capture LEO networks’ dynamic behavior. Inefficient placement increases propagation delays, reduces performance, and wastes resources. Therefore, a dynamic controller assignment strategy is essential to minimize latency and optimize active controllers for energy efficiency.

Dynamic controller placement has been explored in various networks. For instance, software-defined vehicular networks reduce control delay \cite{ref13} but overlook controller overhead, crucial in LEO systems. Switch-to-controller assignments focus on load balancing \cite{ref15} but ignore satellite networks’ rapid topology and traffic changes. Prior SDN-enabled satellite methods adapt to traffic variations considering replacement costs \cite{ref13} or combine static deployment with dynamic switch allocation \cite{ref8}. However, none jointly optimize active controllers and propagation delay in dynamic LEO networks. Recent work \cite{ref16} adjusts controller locations by network load but lacks real-time satellite-to-controller assignment optimization.

Beyond SDN, research improves inter-satellite communication and routing. Federated learning trains distributed AI models in LEO networks via parallel, decentralized schemes to reduce convergence time and energy use \cite{refJ06}. Laser inter-satellite links support adaptive routing to lower latency and increase efficiency \cite{refJ05}. Yet, these approaches neglect the essential satellite-to-controller assignment challenge critical for network performance.

To address this, we formulate the dynamic satellite-to-controller assignment (DSCA) problem as an optimization to minimize propagation delays. We extend it to the optimal dynamic satellite-to-controller assignment (Opt-DSCA), which dynamically pairs satellites with controllers and adjusts the number of active controllers. This balances minimizing active controllers and propagation delays, enhancing network performance, QoS, and communication efficiency while reducing energy use, costs, and management complexity. 

The main contributions of this work are as follows:
\begin{itemize}
\item We have developed an SDN-enabled framework for LEO satellite networks using the OpenFlow protocol, implemented in an OMNeT++-integrated framework, to simulate real-time controller assignment strategies.

\item A DSCA approach is proposed to continuously adjust satellite-controller assignments in response to LEO satellite
movement, minimizing propagation delays.
\item An Opt-DSCA approach is introduced to jointly optimize delay minimization and controller activation, improving overall efficiency. Our results show that Opt-DSCA dynamically activates between 2 and 5 controllers while maintaining propagation delays within an acceptable range (50-65 ms), ensuring both low latency and efficient resource utilization.
\end{itemize}

The rest of the paper is organized as follows. Section \ref{RW} summarizes the existing works on controller assignment problems in satellite networks. Section \ref{SM} describes the SDN-enabled LEO satellite constellation and formulates the DSCA and Opt-DSCA problems into an optimization problem. Section \ref{SDNDI} presents the SDN framework developed and integrated into the OMNeT++ simulator. Section \ref{Result} presents the dynamic assignment results regarding propagation delays. Finally, we conclude the paper in Section \ref{Conclusion}.

\section{Related work} \label{RW}
Controller placement and assignment in SDN have been widely studied in terrestrial and satellite networks. The dynamic controller placement problem in LEO satellite networks is particularly challenging due to rapidly changing topology and strict latency requirements. While terrestrial SDN strategies provide foundational insights, they often fail to adapt to satellite networks' dynamic nature. In terrestrial networks, dynamic placement methods optimize the number and location of active controllers. Bari et al. \cite{ref2} proposed a framework for deploying multiple controllers in WANs to minimize flow setup time and communication costs. He et al. \cite{ref3} formulated a Mixed Integer Programming model to minimize average flow setup time. Dixit et al. \cite{ref6} introduced ElastiCon, a distributed SDN controller architecture that dynamically adjusts pool sizes based on traffic. Wang et al. \cite{ref7} proposed dynamic controller assignment for efficient switch reassignment amid changing conditions.

These terrestrial solutions are less effective for LEO satellite networks due to rapidly fluctuating inter-satellite links and topology shifts, prompting NTN-specific approaches.

In satellite networks, Toufga et al. \cite{ref13} proposed a dynamic controller placement strategy adapting locations based on traffic and replacement costs. Guo et al. \cite{ref8} introduced a hybrid static-dynamic approach combining fixed deployments with dynamic switch allocations. Chen et al. \cite{ref15} developed adaptive controller configurations to balance load and reduce overhead and latency. Other studies explored dynamic controller assignment to optimize switch management. Jiang et al. \cite{ref17} applied a metaheuristic based on the African Vultures Optimization Algorithm for multi-controller placement in SDN-enabled satellites, emphasizing reliability.

Hybrid approaches have also been examined. Reference \cite{ref5} proposed a three-layer hierarchical controller architecture for GEO/LEO SDN networks to minimize interlayer link costs. Liu et al. \cite{ref1} studied joint placement of controllers and gateways to maximize reliability under latency constraints. Torkzaban et al. \cite{ref4} minimized SDN control path failures by deploying controllers near gateways. Another study \cite{ref8} combined static controller placement with dynamic switch assignment.

Despite these efforts, no prior work fully addresses the Dynamic Satellite-to-Controller Assignment (DSCA) problem targeting both propagation delay and active controller count. This study uniquely integrates DSCA within an optimization framework that reduces delay and optimizes controller activation, enabling more efficient and adaptive control in SDN-enabled LEO satellite networks.

\section{SYSTEM MODEL AND PROBLEM FORMULATION} \label{SM}
\subsection{SDN-enabled LEO constellation satellite architecture} \label{AR}
\begin{figure}[htbp]
    \vspace{-1em}  

    \centering
    \includegraphics[width=0.9\linewidth, height=0.17\textheight]{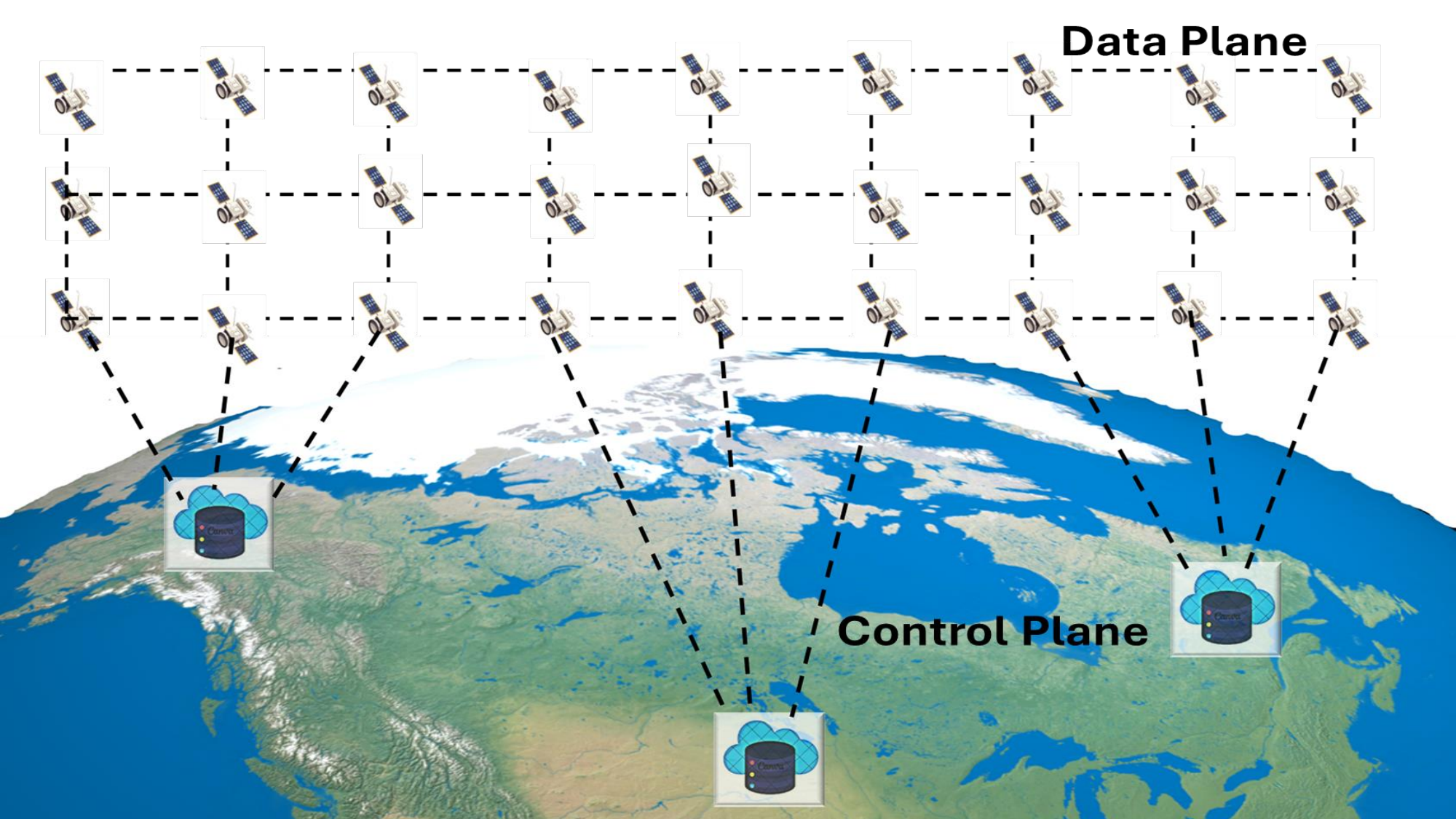}
    \caption{Proposed SDN-enabled LEO constellation architecture.}
    \label{fig: Fig.1}
        \vspace{-1em}  
\end{figure}
The proposed architecture for an SDN-enabled LEO constellation is shown in Fig.\ref{fig: Fig.1}; It comprises two main layers: the data plane layer and the control plane layer. The data plane is primarily composed of the LEO satellites, which form the backbone of the SDN constellation. To ensure continuous global coverage, LEO satellites use two types of inter-satellite links (ISLs): Intra-Orbit ISLs, connecting neighboring satellites within the same orbital plane, and Inter-Orbit ISLs, linking satellites in adjacent orbital planes for cross-orbit communication.
On the other hand, the control plane layer, an integral part of the SDN-enabled LEO constellation, is managed by SDN controllers positioned strategically at fixed locations on Earth. These controllers continuously monitor the network's operation.

In our model, all satellites are SDN switches, and they are assigned to a certain controller; the controller is in charge of them and manages their flow tables. Switches only need to handle data packets following the instructions of their controllers. Consider a set of controllers, denoted as \( c_j \in C \), where the total number of controllers is \( m \). Similarly, let there be a set of satellites, represented by \( s_i \in S \), with a total of \( n \) satellites. We define a binary variable \( x^{[l]}_{i,j} \) to indicate whether satellite \( i \) is assigned to controller \( j \) at time slot \textit{l}, as follows:
\begin{equation}
 \hspace{-.5cm} 
x^{[l]}_{i,j} =
\begin{cases}
1 \text{, if satellite } i \text{ is assigned to controller } j, \\
0 \text{, otherwise.}
\end{cases}  
\end{equation}
An $n \times m$ matrix is defined to present $\mathbf{x}$, where $n$ is the number of satellites and $m$ is the number of controllers.

The activation status of controller $j$ at time slot \textit{l}, with $j \in \{1, 2, \dots, m\}$, is represented by the binary variable $y^{[l]}_j$: 
\begin{equation}
 \hspace{-.5cm} 
y^{[l]}_j =
\begin{cases}
1 \text{, if controller } j \text{ is active,} \\
0 \text{, if controller } j \text{ is inactive.}
\end{cases}
\end{equation}

We can represent the vector of controller activation at time slot \textit{l} as \( y^{[l]} = \begin{bmatrix} y_1 & y_2 & \cdots & y_m \end{bmatrix} \), where each element of the vector corresponds to the activation status of a specific controller.
If satellite $s_i$ is re-assigned from its current controller to a new controller $c_\xi$ at time slot \textit{l}, the assignment matrix $\mathbf{x}$ should be updated as follows:
\begin{itemize}
    \item Set the current satellite $s_i$ to its previous controller to 0:
\begin{equation*}
   x^{[l]}_{i,j} = 0 \quad \text{for the current } c_j \text{ of } s_i \text{ at time slot \textit{l}}. 
\end{equation*}
\item Assign satellite \( s_i \) to controller \( c_\xi \) by setting it to 1:
\begin{equation*}
    x^{[l]}_{i,\xi} = 1 \quad \text{for the new } c_\xi \text{ assigned to } s_i \text{ at time slot \textit{l}}.
\end{equation*}
\end{itemize}
Thus, the assignment matrix $\mathbf{x}$ is updated to reflect the reassignment of $s_i$ from $c_j$ to $c_\xi$ in the time slot \textit{l}.

\subsection{Optimization problem formulation} \label{OPF}
The DSCA and Opt-DSCA approaches are formulated as mixed integer linear programming (MILP) models. The DSCA approach is developed to minimize the average propagation delay under specific constraints, focusing solely on propagation delay while excluding transmission, queuing, and forwarding delays. In contrast, the Opt-DSCA model simultaneously minimizes both the average propagation delay and the number of active controllers, ensuring balanced trade-offs. Both formulations are solved using CPLEX, implemented in Python. This section provides the problem formulation for both DSCA and Opt-DSCA.

\subsubsection{DSCA Problem Formulation}
The DSCA approach aims to minimize the average propagation delay between each controller \(c_j\) and satellite \(s_i\) in the time slot \textit{l}, denoted as \(d^{[l]} (s_i, c_j)\). The shortest path from a satellite to a visible ground station is calculated using Dijkstra's algorithm. This algorithm determines the path with the least delay, considering both intra-satellite and inter-satellite links. It adapts to satellite movement by first identifying the visible ground stations within the satellite's coverage area and then computing the shortest path through the satellite network. Propagation delays are determined through simulations in OMNeT++.

The objective function minimizes the average propagation delay using normalization and weighting for balanced optimization. Equation~\eqref{eq:f1} shows the normalized delay at time slot~$l$, ensuring extreme values do not skew results. The delay is normalized as 
\begin{equation}
   \bar{d^{[l]}}_{\text{norm}}(s_i, c_j) = \frac{d^{[l]}(s_i, c_j) - d_{\min}}{d_{\max} - d_{\min}}, 
\end{equation}
where \( d_{\min} \) and \( d_{\max} \) represent the minimum and maximum propagation delays observed across all pairs and time instances, respectively. This normalization ensures the delays are scaled between 0 and 1, providing comparability across varying ranges of delay values.

The weighted average propagation delay is then defined as
\begin{equation}
    f_1 = \frac{1}{n} \sum_{i=1}^{n} \sum_{j=1}^{m} w_{\text{delay}} \cdot \bar{d^{[l]}}_{\text{norm}}(s_i, c_j) \cdot x^{[l]}_{ij},
    \label{eq:f1}
\end{equation}
where \( w_{\text{delay}} \) is a weight applied to prioritize the reduction of propagation delay in the optimization process. The weight \( w_{\text{delay}} \) allows for adjusting the emphasis on minimizing latency relative to the other objectives.

The objective is then expressed as
\begin{equation}
\text{minimize } 
\left( 
\frac{1}{n} \cdot \sum_{i=1}^{n} \sum_{j=1}^{m} w_{\text{delay}} \cdot \bar{d}^{[l]}_{\text{norm}}(s_i, c_j) \cdot x^{[l]}_{ij}
\right)
\label{eq:objective1}
\end{equation}
\textbf{Subject to:}
\vspace{-7pt}  

\begin{equation}
    \hspace{-4cm} 
    \sum_{c_{j} \in C}  x^{[l]}_{s_i,c_j} = 1, \quad \forall s_i \in S,
    \label{eq:constraint1}
\end{equation}
\vspace{-7pt}  
\begin{equation}
    x^{[l]}_{s_i,c_j} = 
    \begin{cases} 
      0, & \text{if } d^{[l]}(s_i, c_j) > \text{max\_propagation\_delay}, \\
      1, & \text{otherwise},
    \end{cases}
    \label{eq:constraint2}
\end{equation}
\vspace{-7pt}  
\begin{equation}
\hspace{-6.2cm} 
    \sum_{c \in C}  y^{[l]}_{c} = K,
     \label{eq:constraint3}
\end{equation}
\vspace{-7pt}  
\begin{equation}
\hspace{-3.5cm} 
    x^{[l]}_{s_i,c_j} \leq y^{[l]}_{c_j}, \quad \forall s_i \in S, \forall c_j \in C.
     \label{eq:constraint4}
\end{equation}
\vspace{-11pt}  

The constraints governing this dynamic controller assignment problem include:
Equation \eqref{eq:constraint1} guarantees that each LEO satellite is assigned to exactly one controller, clearly defining management responsibilities and preventing any overlap or ambiguity in control assignments. Equation \eqref{eq:constraint2} ensures that the binary variable \(x^{[l]}_{i,j}\) is set to 0 in the time slot \textit{ l} if the average propagation delays between a satellite and its assigned controller exceed the maximum allowable latency. 
Equation \eqref{eq:constraint3} ensures that the total number of controllers to be placed in the network is \textit{K} at time slot \textit{l}. Equation \eqref{eq:constraint4}  ensures that a satellite $s_i$ is controlled by an active controller $c_j$ at time slot \textit{l}.

\subsubsection{Opt-DSCA Problem Formulation}
The Opt-DSCA aims to minimize the average propagation delay and reduce the number of active controllers while satisfying constraints \eqref{eq:constraint1}, \eqref{eq:constraint2}, and \eqref{eq:constraint4}. Constraint \eqref{eq:constraint3} is unnecessary, as the optimal number of active controllers is dictated by the network's specific needs.

The second objective, \( f_2 \), represents the normalized number of active controllers, defined as
\begin{equation}
  f_2 = \frac{(1 - w_{\text{delay}})}{m} \cdot \sum_{j=1}^{m} y^{[l]}_j,  
\end{equation}
where \((1 - w_{\text{delay}})\) is a weighting factor that balances the importance of minimizing the number of active controllers at time slot \textit{l}.

The opt-DSCA approach aims to minimize both the normalized average propagation delay and the number of active controllers. Normalizing propagation delays allows for a fair comparison across different satellite-controller pairs, especially when facing extreme variations. Pairs with significantly larger delays could disproportionately affect the objective function without normalization, leading to suboptimal decisions. The weighting parameters \( w_{\text{delay}} \) and \( (1 - w_{\text{delay}}) \) allow network operators to prioritize either latency reduction or controller minimization, depending on specific network requirements.

Mathematically, the objective is to minimize the weighted sum of the normalized average propagation delay and the proportion of active controllers, expressed as
\begin{equation}
\resizebox{\columnwidth}{!}{%
  $\text{minimize} \; \displaystyle \left( \frac{1}{n} \sum_{i=1}^{n} \sum_{j=1}^{m} w_{\text{delay}} \cdot \bar{d}^{[l]}_{\text{norm}}(s_i, c_j) \cdot x^{[l]}_{ij} + \frac{(1 - w_{\text{delay}})}{m} \sum_{j=1}^{m} y^{[l]}_j \right)$
}
\label{eq:objective2},
\end{equation}
where the first term captures the normalized delay across all satellite-controller assignments, and the second term penalizes the number of active controllers in the network.

\section{SDN Development and Integration in NTN Using OMNeT++}  \label{SDNDI}
In Software-Defined Networking (SDN), separating control and data planes enables dynamic and efficient network management. The \textit{control plane}, managed by SDN controllers, determines forwarding decisions based on network conditions. The \textit{data plane} includes SDN-enabled switches that forward packets per flow table instructions, while the \textit{management plane} supports monitoring and traffic control.

A simulation model for SDN in LEO satellite networks was developed using OMNeT++ and INET \cite{refJ01}, extending the Open Source Satellite Simulator (OS3) \cite{refJ02} with realistic satellite mobility and dynamic routing via Dijkstra’s algorithm.

SDN principles were integrated into NTN to address latency and dynamic topologies. Using OpenFlow, the controller communicates with satellites through a standardized interface, enabling centralized control and adaptive routing. The SDN architecture consists of LEO satellites in the data plane acting as OpenFlow switches that send \textit{Packet-In} messages to the controller via ground stations when no matching rule exists. The control plane updates flow tables through \textit{Flow-Mod} messages, enabling real-time route adjustments. This framework supports scalable, adaptive management of dynamic LEO satellite networks, enhancing responsiveness and performance in NTN environments.

\section{System Performance Evaluation} \label{Result}
In this section, we provide the simulation setting and evaluate the performance of DSCA and Opt-DSCA approaches.

\subsection{Simulation Settings}
We use the OMNeT++ simulator to model and simulate the satellite network, which comprises 66 LEO satellites organized into six orbital planes. The satellite's inclination is set to 98.98 degrees, and its altitude is 1325 kilometers.

In our optimization problem, propagation delay weights 75\%, and the number of active controller weights 25\%. We have prioritized minimizing latency while considering the importance of reducing the number of active controllers in the network with this weight distribution. By emphasizing propagation delay, our goal is to enhance the overall performance of the satellite communication system while maintaining efficient controller utilization.
\subsection{Results}
To evaluate the impact of controller assignment on network performance, we utilized and compared the following three approaches: (1) The static satellite-to-controller assignment (SSCA) approach pre-assigned each LEO satellite to a controller; thus, satellite-to-controller assignments remained static throughout the simulation. (2) The DSCA approach introduced dynamic assignment, allowing LEO satellites to be dynamically reassigned to controllers to reduce the average propagation delay. (3) The Opt-DSCA approach improved further by dynamically assigning LEO satellites while determining the optimal number of active controllers to minimize propagation delays.
In the static assignment approach we used OMNeT++ to simulate the network's topology with fixed controller and satellite assignments. In the dynamic assignment and optimal number of activated controllers strategy we employed the CPLEX optimizer to solve the optimization problem. The dynamic assignment strategy focused on minimizing propagation delays by dynamically reassigning LEO satellites to controllers, while the third approach not only dynamically assigned satellites but also optimized the number of active controllers to reduce delays further.
\begin{figure}[htbp] 
    \vspace{-1em}  

    \centering
    \begin{minipage}{\linewidth}
        \centering
        \includegraphics[width=0.9\linewidth, height=0.18\textheight]{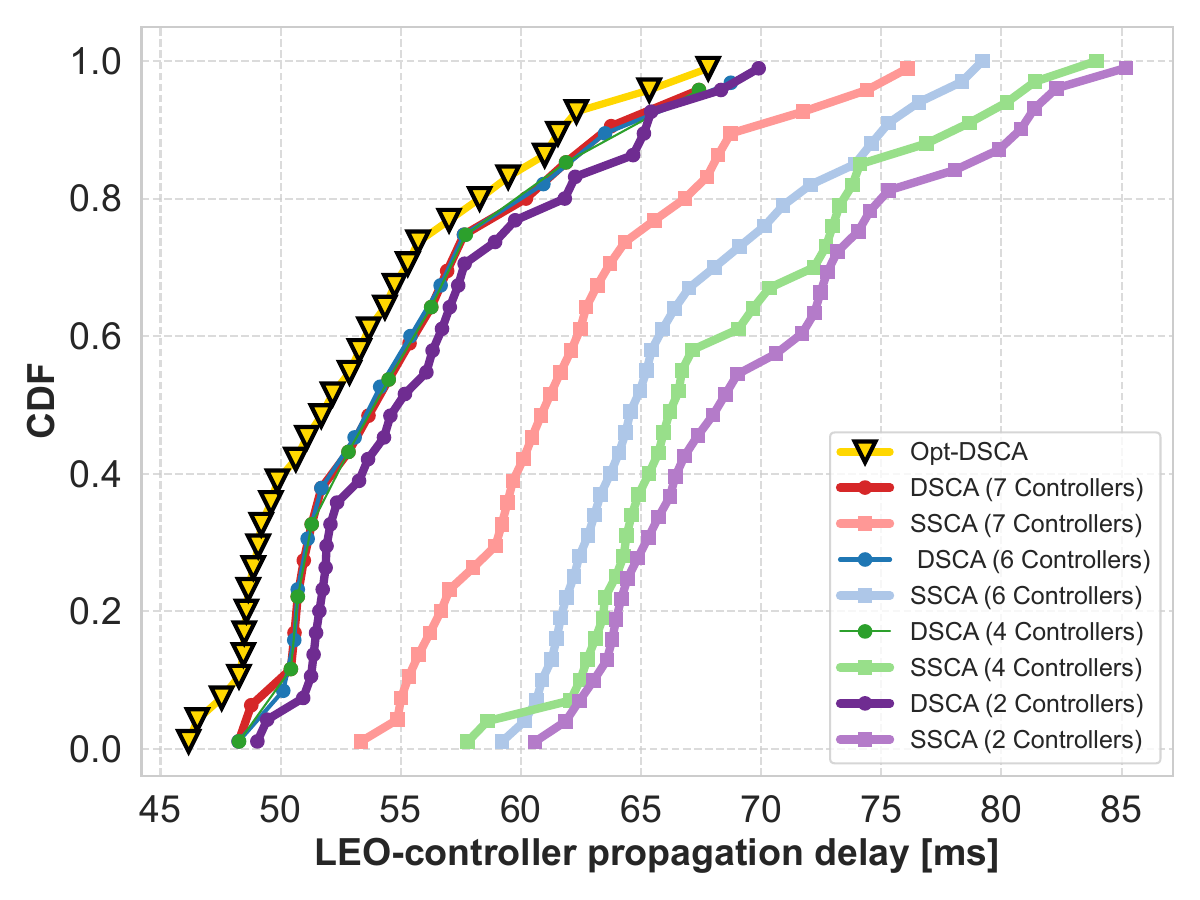} 
        \caption{Comparison of SSCA, DSCA, and Opt-DSCA approaches.}
        \label{fig:Performance analysis}
    \end{minipage}
        \vspace{-1em}  

\end{figure}
Initially, we activated two controllers placed in the most populated cities in Canada: Toronto and Montreal. In subsequent runs, we incrementally increased the number of active controllers by one, placing them based on the population size. Controllers were progressively added in the following cities: Toronto, Montreal, Vancouver, Calgary, Edmonton, Ottawa, and Mississauga. At each step, we simulated the network with an increasing number of controllers, up to a maximum of seven.

Figure \ref{fig:Performance analysis} presents a cumulative distribution function (CDF) analysis comparing the performance of three controller assignment approaches, Opt-DSCA, DSCA, and SSCA, when applied to a satellite network regarding propagation delays. The x-axis represents propagation delays in milliseconds, while the y-axis shows the CDF, from 0 to 1, showing the proportion of communication instances below a given delay threshold.
The SSCA approach results exhibit higher propagation delays, particularly with fewer controllers. This is especially evident with two controllers, where a significant proportion of satellites experience delays above 70 ms, as indicated by the shift to the right of the light purple curve. This is a consequence of SSCA’s inability to adapt to the dynamic environment of LEO satellites, whose rapid movement frequently changes their relative positions to controllers. The static nature of SSCA produces highly suboptimal communication paths, increasing latency by assigning satellites to non-ideal controllers according to distance. As the number of controllers increases, a shift to the left can be observed for the CDF curves, representing reduced delays. Beyond seven controllers, improvement diminishes as far as performance gains flatten. For example, around 70\% of the communications with seven controllers had delays below 65 ms, while with two controllers the delay was closer to 70 ms. 

\begin{figure} [htbp]
          \vspace{-1em}  
    \centering
    \begin{minipage}{\linewidth}
        \centering
        \includegraphics[width=0.9\linewidth, height=0.18\textheight]{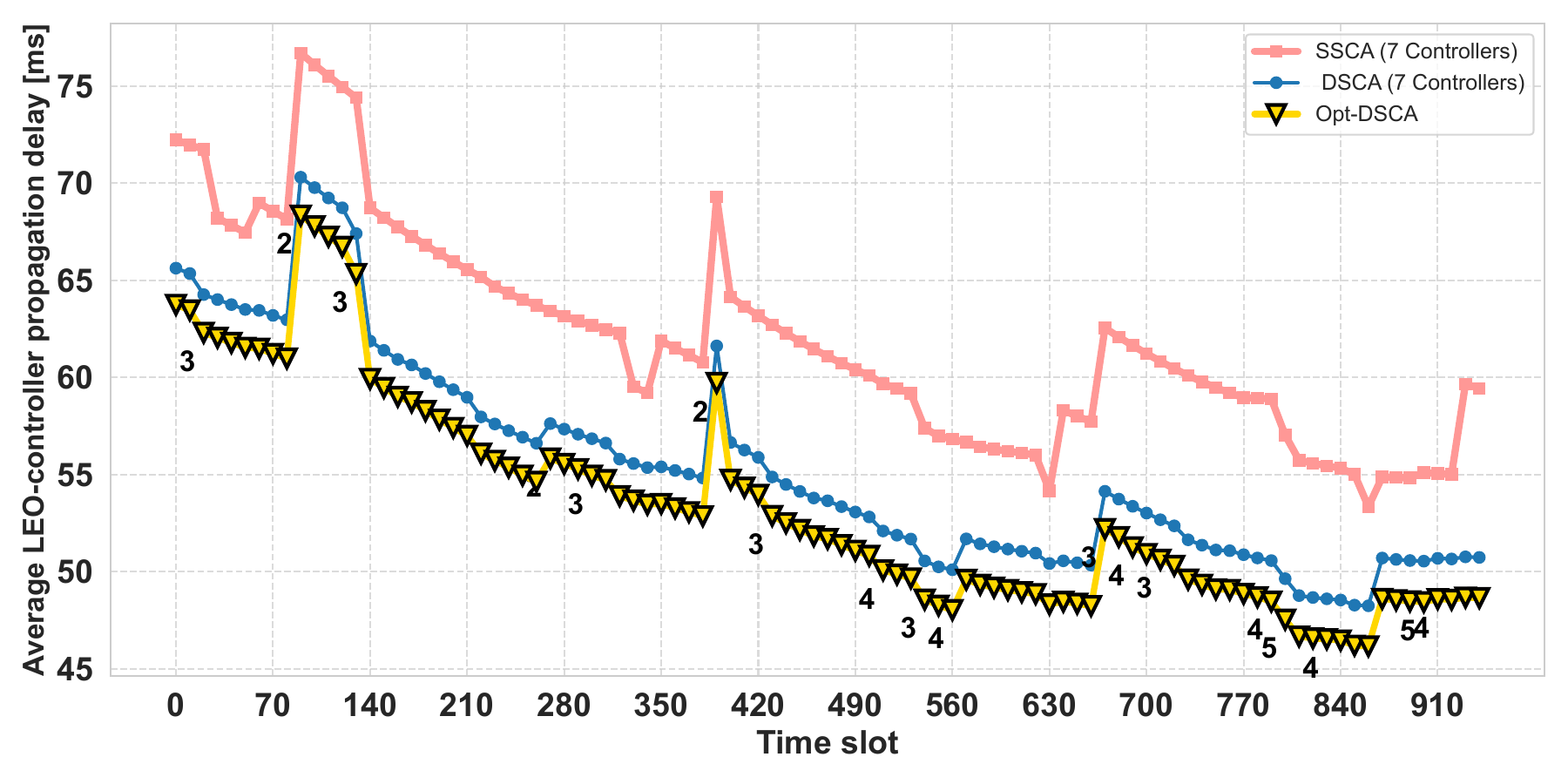} 
        \caption{Variation in active controllers and propagation delay for Opt-DSCA (Optimal) vs. SSCA and DSCA (Fixed 7 controllers).}
        \label{fig:NumberCntrs}
    \end{minipage}
    \vspace{-1em}  
\end{figure}

On the other hand, the DSCA approach performs more efficiently because of the dynamic assignment of LEO satellites to controllers. Here, DSCA has shorter communication paths at every moment by continuously evaluating propagation delays and adapting to satellite movement. The CDF curves for DSCA reflect this, with the scenario using seven controllers (dark red curve) achieving the lowest propagation delays. As the number of controllers decreases, the curves shift to the right, indicating increasing delays. Even with four controllers, the DSCA outperforms the SSCA. At approximately 55 ms, around 60\% of communications experience delays with DSCA, whereas with SSCA, delays are observed at around 70 ms. This highlights DSCA’s superior efficiency in managing the dynamic nature of LEO satellite networks, though the marginal benefit of adding controllers beyond four diminishes. When comparing Opt-DSCA approach to both DSCA and SSCA approaches, the advantage of adaptability becomes clear. Opt-DSCA approach, which dynamically adjusts the number of active controllers based on the network's needs, achieves the lowest propagation delays, as indicated by its curve being furthest to the left in the figure. Even with fewer active controllers, the Opt-DSCA approach manages to maintain lower delays compared to both DSCA and SSCA approaches, highlighting the importance of flexibility in controller management. DSCA approach, though superior to the SSCA approach, lacks Opt-DSCA’s ability to flexibly adjust the number of active controllers, resulting in higher delays, while the SSCA approach performs the worst overall due to its static nature.

Figure \ref{fig:NumberCntrs} compares the propagation delays of SSCA, DSCA, and Opt-DSCA approaches over time. SSCA, with its fixed use of seven controllers, maintains a stable but relatively high delay and cannot adapt to changing network conditions over time. This demonstrates the limitations of static approaches in dynamic environments. Alternatively, DSCA adjusts satellite assignments based on real-time conditions but is limited by a fixed seven-controller. While the dynamic assignment of satellites reduces delays to some extent, its performance is still limited by the maximum controller threshold. The Opt-DSCA approach minimizes propagation delays by dynamically assigning controllers per time slot, optimizing resource use. The figure annotations show the number of active controllers when updates occur. For example, at time slot 420, only three controllers are active, leading to lower delays compared to SSCA and DSCA. These controllers remain active until time slot 500, when four are activated. This showcases the effectiveness of dynamic and optimized controller allocation, demonstrating that fewer controllers can outperform when resources are allocated based on real-time conditions.

\begin{figure}[htbp]  
    \vspace{-1em}  
    \captionsetup{skip=0pt}  
    \begin{subfigure}{\linewidth}
        \centering
        \includegraphics[width=\linewidth, height=0.18\textheight]{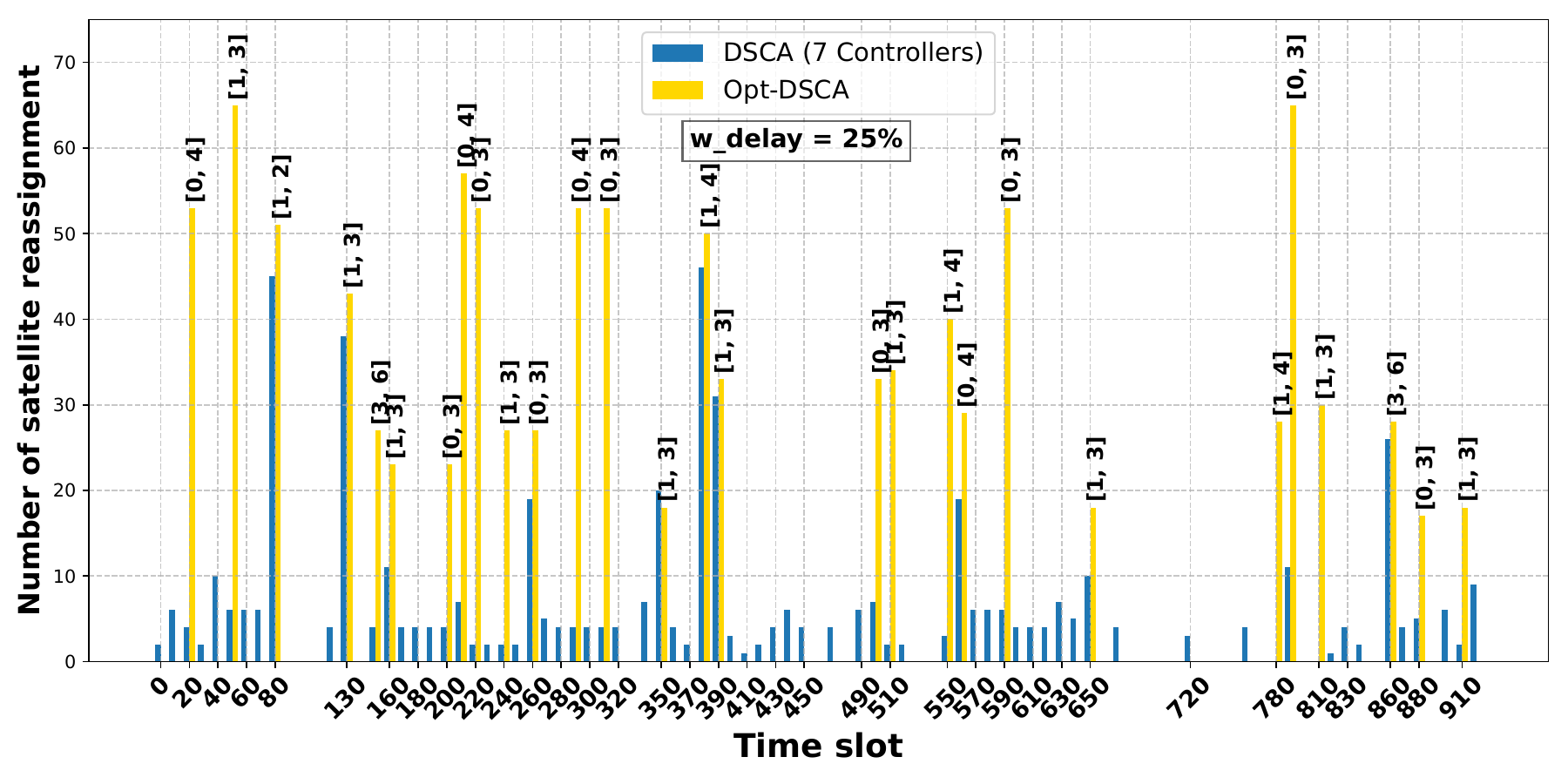}
        \label{fig:comparison_25}
    \end{subfigure}
    \par\vspace{-10pt}
    \begin{subfigure}{\linewidth}
        \centering
        \includegraphics[width=\linewidth, height=0.18\textheight]{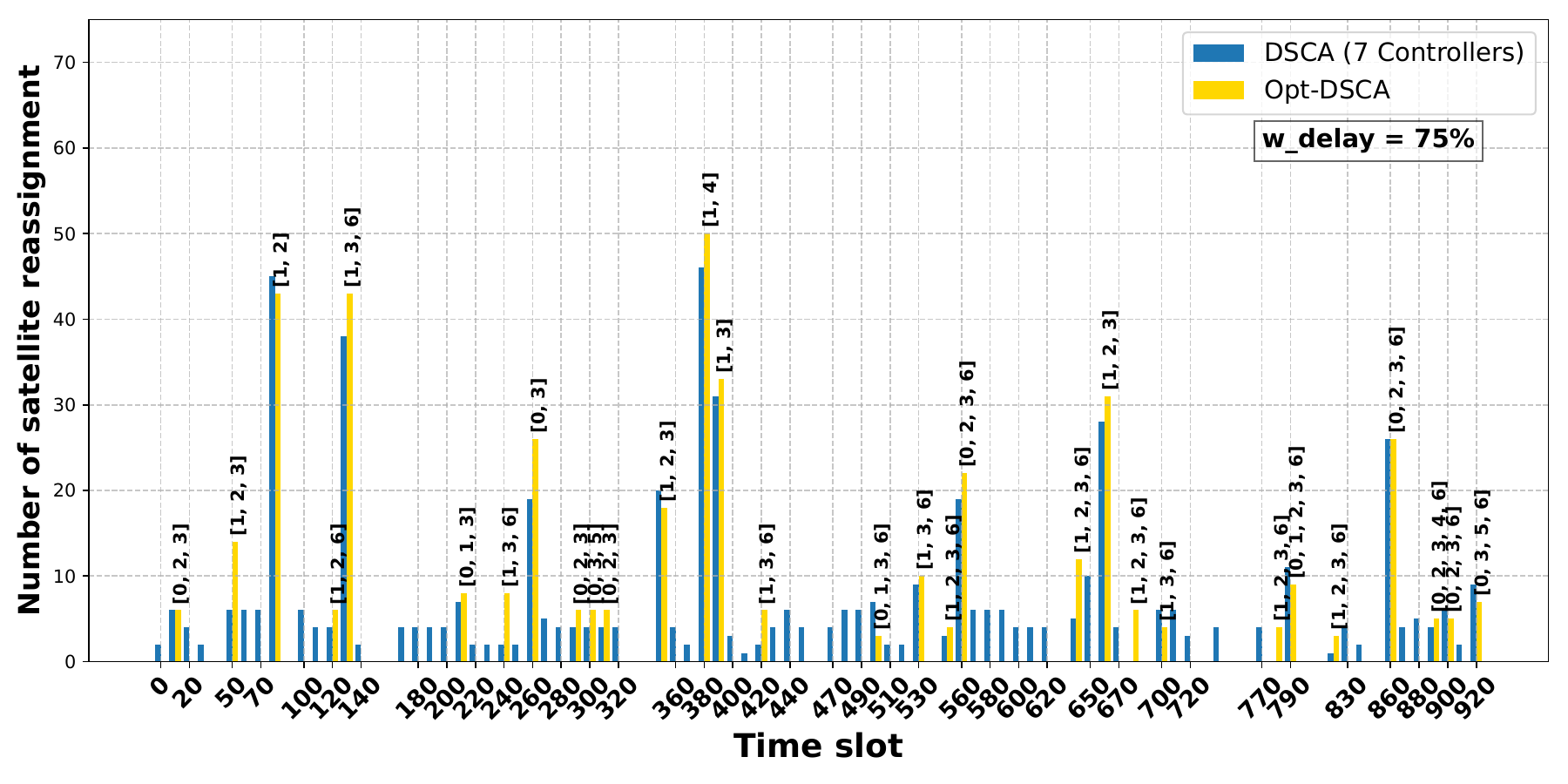}
        \label{fig:comparison_75}
        
    \end{subfigure}
     \par\vspace{-10pt}
    \caption{Comparison of satellite reassignment frequency in DSCA and Opt-DSCA approaches.}
        \label{fig:comparison}
          \vspace{-1em}  
\end{figure}
Figure \ref{fig:comparison} illustrates the frequency of satellite reassignments across different time slots for DSCA, which operates with a fixed 7-controller setup, and Opt-DSCA, which dynamically adjusts the number of active controllers. The annotations in the figure indicate the active controllers at each time slot where updates occur. For instance, in the first subfigure in time slot 80, the annotation "[1, 2]" signifies that the controllers with indices 1 and 2 are active. If no annotation appears for a particular time slot, it implies that no changes have been made to the active controllers during that time. The two subfigures correspond to different values of \textit{$w\_delay$} (25\% and 75\%), representing the trade-off between minimizing propagation delay and optimizing controller activation. DSCA consistently exhibits higher reassignment frequencies due to its inflexible approach of keeping all seven controllers active at all times. In contrast, Opt-DSCA demonstrates an adaptive strategy by dynamically activating controllers, thereby reducing unnecessary reassignments as \textit{$w\_delay$} increases.

With low \textit{$w\_delay$} (25\%), the optimization prioritizes minimizing the number of active controllers, leading to fewer active controllers per time slot but more frequent satellite reassignments (yellow bars) as satellites are redistributed among a limited set of controllers. At high \textit{$w\_delay$}  (75\%), Opt-DSCA focuses more on minimizing propagation delay by allowing more controllers to be active when needed, leading to fewer satellite reassignments as satellites remain more stable in their assignments. This results in a significant reduction in reassignment events compared to lower \textit{$w\_delay$} scenarios while still ensuring that delay is minimized. Opt-DSCA activates controllers dynamically while maintaining lower delays, unlike DSCA, which inefficiently keeps all seven controllers active at all times. Despite the presence of blue bars, the absence of yellow bars at specific time slots indicates that Opt-DSCA avoided unnecessary reassignments, highlighting its efficiency.

In summary, dynamic controller activation is advantageous for LEO satellite networks enabled with SDN. In comparison to both SSCA and DSCA, Opt-DSCA provides an effective balance between propagation delay minimization and resource optimization. Due to its static nature, SSCA suffers from high propagation delays, whereas DSCA improves performance by dynamically reassigning satellites, but remains constrained by its fixed controller setup. In contrast, Opt-DSCA achieves the lowest propagation delays by intelligently adjusting the number of active controllers, ensuring an efficient and scalable solution for satellite network management.
\section{Conclusions} \label{Conclusion}
Our analysis confirms that dynamic controller assignment significantly enhances network performance in SDN-enabled LEO satellite constellations. The comparison of SSCA, DSCA, and Opt-DSCA approaches highlights that static controller assignment results in higher propagation delays, especially with limited controllers. In DSCA, satellites are dynamically reassigned, reducing delay but maintaining a fixed number of controllers. Opt-DSCA, on the other hand, optimizes controller activation based on real-time network conditions, resulting in the lowest delays as well as efficient resource utilization.
The results underscore the benefits of adaptability in controller assignment, with Opt-DSCA consistently outperforming other approaches in minimizing delays and optimizing network efficiency. The CDF analysis demonstrates how Opt-DSCA maintains lower delays even with fewer active controllers, proving its effectiveness in balancing controller activation and satellite reassignment.
These findings reinforce the necessity of flexible and intelligent controller activation strategies in LEO satellite networks. Opt-DSCA presents a scalable and resource-efficient solution for reducing propagation delays while maintaining network stability. In summary, this research provides further insight into controller assignments toward optimization within SDN-enabled LEO satellite networks and opens up avenues for future research into satellite communications-integrated terrestrial networks.

\section*{Acknowledgment}
This work has been supported by the National Research Council Canada, MDA, Mitacs, and Defence R\&D Canada, within the Optical Satellite Communications Consortium framework in response to the High Throughput Secure Networks challenge program of the Government of Canada.

\vspace{12pt}
\end{document}